\documentclass[twocolumn,superscriptaddress]{revtex4-1}

\usepackage{amsmath}
\usepackage{amssymb}
\usepackage{braket}
\usepackage{mathtools}

\DeclarePairedDelimiter\floor{\lfloor}{\rfloor}
\usepackage{pgfplots}

\begin{document}

\title{Current quantization and fractal hierarchy in a driven repulsive lattice gas}

\author{Pietro Rotondo}
\affiliation{School of Physics and Astronomy, University of Nottingham, Nottingham, NG7 2RD, UK}
\affiliation{Centre for the Mathematics and Theoretical Physics of Quantum Non-equilibrium Systems,
University of Nottingham, Nottingham NG7 2RD, UK}
\author{Alessandro Luigi Sellerio}
\affiliation{Dipartimento di Fisica, Universit\`a degli Studi di Milano, via Celoria 16, 20133 Milano, Italy}
\author{Pietro Glorioso}
\affiliation{Dipartimento di Fisica, Universit\`a degli Studi di Milano, via Celoria 16, 20133 Milano, Italy}
\author{Sergio Caracciolo}
\affiliation{Dipartimento di Fisica, Universit\`a degli Studi di Milano, via Celoria 16, 20133 Milano, Italy}
\affiliation{INFN Milano, via Celoria 16, 20133 Milano, Italy}
\author{Marco Cosentino Lagomarsino}
\affiliation{Sorbonne Universit\'es, UPMC Univ Paris 06, UMR 7238, Computational and
Quantitative Biology, 5 Place Jussieu 75005 Paris, France}
\affiliation{CNRS, UMR 7238, Computational and
Quantitative Biology,  France}
\author{Marco Gherardi}
\affiliation{Dipartimento di Fisica, Universit\`a degli Studi di Milano, via Celoria 16, 20133 Milano, Italy}
\affiliation{Sorbonne Universit\'es, UPMC Univ Paris 06, UMR 7238, Computational and
Quantitative Biology, 5 Place Jussieu 75005 Paris, France}

\begin{abstract}
  Driven lattice gases are widely regarded as the paradigm of
  collective phenomena out of equilibrium.  While such models are
  usually studied with nearest-neighbor interactions, many empirical
  driven systems are dominated by slowly decaying interactions such as
  dipole-dipole and Van der Waals forces.  Motivated by this gap, we
  study the non-equilibrium stationary state of a driven lattice gas
  with slow-decayed repulsive interactions at zero temperature.
  By numerical and analytical calculations of the particle current as
  a function of the density and of the driving field, we identify (i)
  an abrupt breakdown transition between insulating and conducting
  states, (ii) current quantization into discrete phases where a
  finite current flows with infinite differential resistivity, and
  (iii) a fractal hierarchy of excitations, related to the Farey sequences of number theory.
  We argue that the origin of these effects is the competition between
  scales, which also causes the counterintuitive phenomenon that
  crystalline states can melt by increasing the density.
\end{abstract}

\pacs{}
\maketitle

\section{Introduction}

In some areas of statistical physics --- most notably for
ferromagnetism, where the Ising model is the paradigmatic framework
---
classical coarse-grained theories have led to notable conceptual
advances.
For non-equilibrium transport, the reference models fall in the class
of driven lattice gases (DLG), which were originally introduced to
capture collective phenomena out of equilibrium~\cite{Katz:1984}.
Such driven diffusive systems play an important role in
non-equilibrium statistical physics~\cite{SchmittmannZia:1995}, as
well as providing models for various transport processes ranging from
vehicular traffic~\cite{Chowdhury2000} to biological
transport~\cite{Chowdhury2004,Chowdhury2005,ParmeggianiFranosch:2004,Parmeggiani2004,Parmeggiani2003,Lipowsky2001}.
At variance with equilibrium lattice gases, the DLG involves a driving
field $E$, that causes particles to hop preferentially in one direction.
The resulting steady states are characterized,
for non-conservative fields (e.g., under periodic boundary conditions),
by a net macroscopic current, i.e., they are genuinely non-equilibrium steady states.
These models show a rich phenomenology, including non-equilibrium phase
transitions and ordering phenomena such as pattern formation, 
self-organization, and morphogenesis.

In DLGs, the attention is usually restricted to nearest-neighbor
Hamiltonians~\cite{BeckerNelissen:2013, TailleurKurchan:2007,
  ParmeggianiFranosch:2004, CaraccioloGambassi:2003,
  delosSantosMunoz:2001}.  However, in certain driven systems,
longer-range interactions are an important defining ingredient. 
We particuarly think of the context of hard condensed matter and
ultracold atoms~\cite{LeviLesanovsky:2014, OlmosYu:2013,
  Rademaker:2013, SaluskaKotur:2006}, where slow-decaying (especially
repulsive) interactions are widespread, for instance, in dipolar
fermions and Rydberg gases, where dipole-dipole and Van der Waals
forces dominate (e.g., the dynamical crystallization of a 1d lattice gas
made of Rydberg atoms has been observed
experimentally~\cite{Schauss:Science:2015}).
In this area, repulsive long-range lattice gases at equilibrium are
sometimes used as reference models.  In particular, they are invoked
to explain the interesting phenomenology, such as devil's staircases
and commensurability transitions, arising in experimental setups where
a lattice spacing competes with the typical inter-particle distance at
fixed filling fraction \cite{ZhangFan:2015,LauerMuth:2012}.
These models (such as those introduced by Frenkel and Kontorova,
Hubbard, and Bak) have a long history and a prominent place in the
statistical mechanics literature, where they are regarded as
prototypes of systems with competing
interactions~\cite{Hubbard:PRB:1978,Bak:PRL:1982,Bak:RPP:1982,LowEmery:1994,BraunKivshar:BOOK,Selke:1988}. 
Despite of the great interest met by the theoretical investigations of
these models at equilibrium, explorations of their properties far from
equilibrium are lacking.
Notably, many charge transport phenomena in condensed matter are still
debated or unexplained theoretically, highlighting the need for the
study of such paradigmatic tractable models.
Particularly challenging is the highly non-linear behavior in current
flow that signals anomalous transport in strongly correlated quantum
many-body systems, which gives rise to complex phase
diagrams~\cite{Dagotto:2005}.
Striking examples of electronic phases with anomalous emergent
transport are the fractional quantum Hall effect (FQHE), the staircase
of fixed-current phases in low-dimensional charge-density wave
conductors (due to phase locking under AC bias voltage), resistive
switching in Mott insulators, and the anomalous current-voltage
characteristics in disordered films~\cite{Nakamura:2013, Zybtsev:2010,
  CarioVaju:2010, Vinokur:2008,Altshuler:PRL:2009}.
%
%

Here, we define and explore a one-dimensional DLG model with
slowly decaying repulsive interactions.  We fully characterize the
behavior of the macroscopic current at varying field and density, by both direct simulation and
analytical calculations.
As we will show, this simple statistical mechanical model exhibits a
rich phenomenology of anomalous transport, resembling some aspects of
strongly interacting many-body systems out of equilibrium.
%


\section{Model}
The system has $L$ sites and $N$ particles, fixed density $\rho=N/L$,
and periodic boundary conditions. A configuration is specified by the
set of occupation numbers $n_i\in\{1,0\}$ ($i=1,\dots,L$), with
$n_i=1$ if the $i$-th site is occupied and $n_i=0$ otherwise. 
At equilibrium, the particles hop to nearest-neighbour sites
randomly with rate
$w(\Delta H)=\text{min}\left\{1,e^{-\beta \Delta H} \right\}$, where
$\Delta H$ is the change in the energy function
\begin{equation}
H = \sum_{i\neq j} V(|i-j|)n_i n_j
\label{eqenergy}
\end{equation} 
due to the proposed jump, and $\beta$ is the inverse temperature.
$V(x)$ is a repulsive convex potential; in particular dipolar and Van
der Waals interactions are characterized by a power law decay
$V(x)\sim 1/x^{\alpha}$ with exponent $\alpha=3$ and $\alpha=6$
respectively.  In the following, we discuss results obtained with
$\alpha=3$.  However, the phenomenology is
robust, and does not depend on the exact functional form
of the potential, provided that it is convex and vanishes at infinite distances
\cite{Hubbard:PRB:1978}.

Applying a constant field $E$ drives the system out of equilibrium.
Owing to the periodic boundary conditions, $E$ is non conservative,
thus breaking detailed balance 
\cite{SchmittmannZia:1995}.
The microscopic effect of the drive is to bias jumps in one direction,
as reflected by the modified hopping rates
\begin{equation}
  w(\Delta H +\ell E )=\text{min}\left\{1,e^{-\beta (\Delta H+\ell E)} \right\}\,,
\label{eqrate}
\end{equation}    
which take into account the work $\ell E$ against the field.
Here, $\ell = -1$ and $\ell = +1$
correspond respectively to jumps along and opposite to $E$. 
We focus here on the zero temperature case, where the rate is
$w=0$ if $\Delta H+\ell E>0$ and $w=1$ otherwise.

\begin{figure}[t]
\includegraphics[width=0.48 \textwidth]{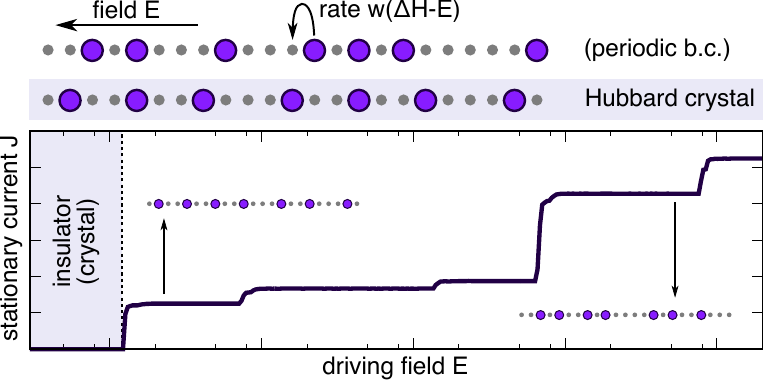}
\caption{ Top: sketch of the model. Mutually exclusive particles hop
  to neighboring sites of a 1D lattice with periodic boundary
  conditions, with rates given by Eq.~(\ref{eqrate}). Particles also
  feel pairwise slow-decayed repulsive interactions.  Bottom: the
  stationary current (here at density $\rho=3/10$) as a function of
  the driving field is zero below a threshold, where the system is
  frozen in the Hubbard state, and anomalously quantized above.  The
  arrows point to typical configurations at the corresponding $E$.  }
\label{figure:picture}
\end{figure}

Let us first briefly consider the zero-field stochastic dynamics,
whereby energy is a non-increasing function of time.
At fixed density $\rho$, the configurations minimizing the energy are
crystalline states with inter-particle distances $\approx1/\rho$.
However, the lattice introduces frustration (at densities different
from $1/q$, with $q \in \mathbb{N}$), forcing some of the
inter-particle distances to deviate from the average.
In the ground state, described by Hubbard
\cite{Hubbard:PRB:1978},
the $n$-th particle occupies the position 
\begin{equation}
x_n=\floor{n/\rho},
\end{equation}
where $\floor{\cdot}$ denotes the integer part.
A useful property of this state is the approximate equidistance
of all $k$-th nearest neighbors.
More precisely, $x_{n+k}-x_n$ is either equal to $r_k$ or to $r_k+1$,
where $r_k=\floor{k/\rho}$, for all $k>0$.
Note that a Hubbard state at density $p/q$, with $p$ and $q$ co-prime,
has period $q$.  Hence, it can only be constructed if the lattice size
is a multiple of $q$.  If this is the case, we refer to the state as
\emph{supported} by the lattice.  The lattice size is set to $L=2520$,
unless specified otherwise (this is the smallest size that supports
all denominators up to $q=10$).
Since the dynamics at $E=0$ is local,
it is not a priori evident that the the system should
always reach the ground state from any starting configuration.
We verified by extensive simulations
that a disordered initial condition (a \emph{quench} from temperature
$T=\infty$) always evolves into the Hubbard crystal in a finite
system.
An example of a Hubbard ground state is shown in Fig.~\ref{figure:picture}
for $\rho=3/10$.

We now turn to the non-equilibrium case $E \neq 0$,
by simulation of the stochastic dynamics (\ref{eqrate}).
If $N$ is the number of particles in the system,
a time step ($t\mapsto t+1$) is realized by $N$ sequential
updates. Each update consists of the choice of a random particle
(uniformly on all particles) and a random neighboring site;
if the site is empty, the particle is moved with probability
$w$ given by (\ref{eqrate}).
Here, the initial condition $\{x_i(0)\}$ is the Hubbard state, but we
tested that a ``hot start'' at infinite temperature 
does not affect the results. 
The driving field is directed towards decreasing positions.
The local current $j_{i+1,i}(t)$ is defined as the particle current
across the bond $(i+1,i)$ between times $t$ and $t+1$, i.e., the
number of particles jumping from site $i+1$ to site $i$ minus the
number of particles jumping in the reverse direction.
We study the behavior of the average stationary current, measured as
\begin{equation}
J = \frac{1}{(T-T_0)L}\sum_{t=T_0}^T \sum_{i=1}^L j_{i+1,i} (t)\,, 
\end{equation}
where $T_0$ is a relaxation time
(fixed \emph{a posteriori} to 100 times the fitted exponential
autocorrelation time of 
$\sum_i j_{i+1,i}$) and $T=10^6$.


\section{Results}

\subsection*{Anomalous quantization of the macroscopic current}
Fig.~\ref{figure:picture} shows the current $J$ as a function of the
applied driving field (here for density $\rho=3/10$).  
At small driving fields, the system
behaves as an insulator with zero stationary current, lying in the
Hubbard ground state.  At a threshold field $E_\mathrm{thr}$ the
insulator breaks down to a conducting state.
A similar (but continuous) transition was found in a simple exclusion process
with next-to-nearest-neighbour interactions at fixed density $1/2$
\cite{Helbing:PRL:1999};
here, owing to the longer range of the interaction, we find transitions for all densities.
In the conducting phase $E>E_\mathrm{thr}$, transport is
anomalous, in that the system preferentially supports a discrete spectrum
of currents.
Rapid changes in current and small near-Ohmic regions alternate with large
plateaux where the differential conductivity
$\sigma(E)=\mathrm{d}J/\mathrm{d}E$ is zero, and the current is
independent of the driving field.

\begin{figure}[t]
\includegraphics[width=0.44 \textwidth]{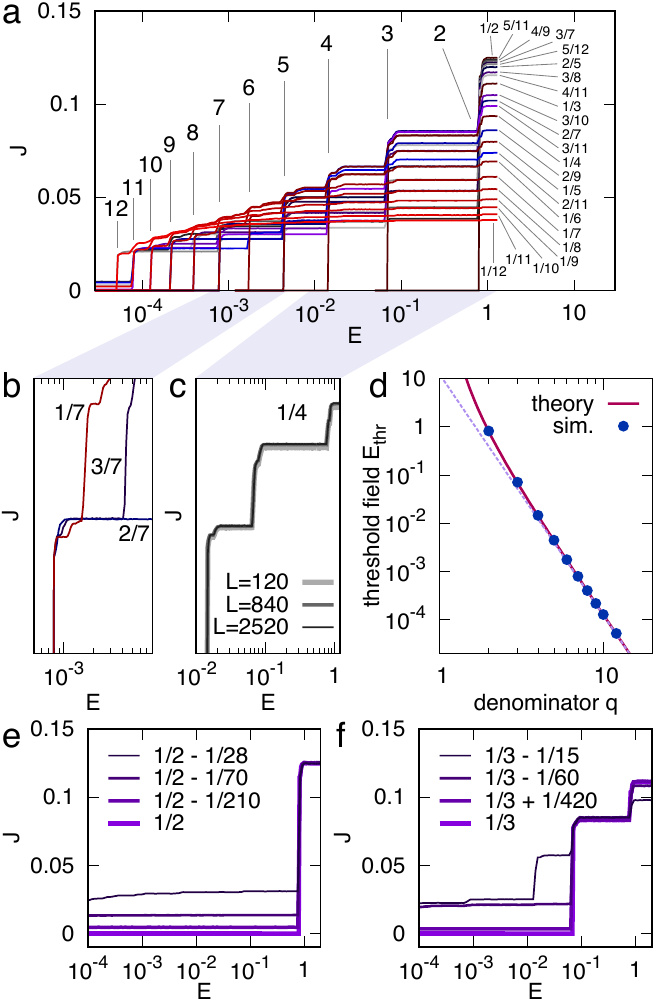}
\caption{
Transitions arise at densities $p/q$,
at threshold fields that only depend on the denominator.
(a) Current versus driving field for various densities $p/q$ (numbers
over the curves are the denominators $q$, with lines pinpointing the
threshold fields). 
(b) A close-up showing
only the curves for $q=7$. 
(c) The current at $\rho=1/4$ for increasing lattice sizes.
(d) How the threshold field depends on the denominator: the circles are
simulations, the solid line is the analytical formula
Eq.~(\ref{eqEthrPwl}), the dotted line is its asymptotic form $q^{-5} \left| \psi_4(1) \right|/2$.
(e,f) Continuity of the current under variation of $\rho$,
around $1/2$ in panel (e) (lines correspond to $\rho=13/28, 17/35, 52/105, 1/2$ from top to bottom)
and around $1/3$ in panel (f) (lines correspond to $\rho=4/15,19/60,47/140,1/3$ from top to bottom).
Errors are smaller than the line widths.
}
\label{figure:JvsE}
\end{figure}

The robustness of this phenomenology is visible in
Fig.~\ref{figure:JvsE}, which shows $J(E)$ for various
densities $\rho\leq 1/2$.  For values of $\rho$ above $1/2$, the plots
are the same as those at $1-\rho$ because of the particle-hole
symmetry $\left\{n_i\mapsto 1-n_i,E \mapsto -E\right\}$.
The behavior of the current is robust in the sense that its features
are independent of several details;
in particular, Fig.~\ref{figure:JvsE} illustrates the following results.
(\emph{i})
The breakdown field $E_\mathrm{thr}(\rho)$ only depends on the
denominator $q$ of the density $\rho$, as we prove theoretically below.
(\emph{ii})
The current does not depend strongly on lattice size; in fact, the
curves for $L=2520$ lie within the error bars of those at $L=840$.
(\emph{iii}) The locations of the main transitions between plateaux
are largely independent of the density.  The current at a given
density $p/q$ appears to transition preferentially at or close to the
threshold fields $E_\mathrm{thr}(1/q')$ of smaller denominators $q'<q$.
This is surprising, as the current-carrying excitations are
expected to be different for different denominators.  
However, this property is a necessary requirement
if the phenomenology is to be stable under small density changes, for
instance when approximating $\rho=p/q$ on a lattice that does not
support it.
As an example, consider the currents for $q=11$.  This denominator is
not supported by $L=2520$, and $\rho=1/11\approx 0.09091$ is rounded
down to $229/2520\approx 0.09087$.  Despite the very different
denominator, the current remains small below $E_\mathrm{thr}(1/11)$ and
becomes larger above.
(iv) The current at the plateaux appears to be a continuous function of the density
at fixed $E$. We checked this by testing several sequences of fractions
converging to low-denominator ones, such as 1/2 and 1/3 (see Fig.~\ref{figure:JvsE}ef for two examples).
Notice that a consequence of property (\emph{iii}) above is that
higher denominators entail curves with larger numbers of plateaux.
Therefore, when considering a sequence with large denominators that converges to a simpler density,
several consecutive plateaux must level out towards the same value.
This behavior is evident, e.g., in Fig.~\ref{figure:JvsE}f.
An important consequence of this continuous behavior is the fact that
the extension of $J(E)$ to irrational values of $\rho$ is unique:
it is possible to approximate the current-field curves
for irrational densities with arbitrary precision in the thermodynamic limit $L\to\infty$.

Altogether, the foregoing observations show that the anomalous
properties of $J(E)$ are not merely reflecting the microscopic details
of the model, but realize a meaningful macroscopic phenomenology.
Furthermore, they suggest that a continuum limit may be definable.

The breakdown field $E_\mathrm{thr}(\rho)$ can be evaluated
analytically as follows.  Its value must be equal to the energy
difference between the crystalline ground state and the first excited
state with a single defect.
Let us consider the ground state $x_n=\floor{n q/p}$ (with $p$ and $q$
coprime), and choose the $n$-th particle, with $n$ a multiple of $p$.
The $k$-th nearest neighbor on its left (right) will be a distance $y_k^\mathrm{L}$ ($y_k^\mathrm{R}$) from $x_n$.
Moving the $n$-th particle along the field (i.e., to the left) will
change the interaction energy by a quantity
$\Delta H = \sum_k \Delta H(k)$, where
\begin{equation}
\label{eq:Hofk}
\Delta H(k)=V(y_k^\mathrm{L}-1)+V(y_k^\mathrm{R}+1)-V(y_k^\mathrm{L})-V(y_k^\mathrm{R})
\end{equation}
is the change of energy due to the interactions with the two $k$-th
nearest neighbors.
Since the state is periodic with period $q$, i.e., it repeats after
$p$ particles, $y_k^\mathrm{L}=y_k^\mathrm{R}=mq$ whenever $k=mp$,
$m\in\mathbb{N}$.  The particles for which this holds will be called
\emph{images} of the one at $x_n$.  Moving the $n$-th particle to the
left changes the interaction energy with its images by a quantity
$\sum_m\Delta H(mp)$ that is computed below.  Let us now focus on the
particles between the first images, thus fixing $1\leq k < p$. 
One has 
\begin{equation}
\begin{split}
x_{n-k}&=\floor{(n-k)q/p}=nq/p-\floor{kq/p}-1\\
x_{n+k}&=\floor{(n+k)q/p}=nq/p+\floor{kq/p};
\end{split}
\end{equation}
these relations are consequences of the fact that $n$ is a multiple of $p$,
that $q$ and $p$ are coprime, and that $k$ is not a multiple of $p$.
Hence, $y_k^\mathrm{L}=y_k^\mathrm{R}+1$;
moreover, it is straightforward to check that this relation holds also if $k>p$,
provided that again $k$ is not a multiple of $p$).  
Therefore, from Eq.~(\ref{eq:Hofk}), $\Delta H(k\neq mp)=0$, and finally
\begin{equation}
\Delta H=\sum_m\Delta H(mp).
\end{equation}
This energy variation holds if $n$ is a multiple of $p$,
but it is the minimum over all particles.
In fact, for all $n$ the image terms are the same as above,
but all other terms in $\Delta H$ are non-negative,
since $x_{n+k}-x_n$ is either $r_k$ or $r_k+1$.
Finally, the energy gap is
\begin{equation}
E_{\mathrm{thr}}(p/q) = \sum_{k=1}^{\infty} \left[V(qk+1)+V(qk-1)-2V(qk)\right].
\label{eqEthr}
\end{equation}
For the power-law potential $V(x)=1/|x|^{\alpha}$ with integer $\alpha$, it evaluates to
\begin{equation}
\begin{split}
E_{\mathrm{thr}}(p/q) = &\frac{(-1)^\alpha}{(\alpha-1)! \, q^\alpha} \left[
\psi_{\alpha-1}(1+q^{-1})    \right. \\
& \left. +\psi_{\alpha-1}(1-q^{-1})-2\psi_{\alpha-1}(1) \right],
\label{eqEthrPwl}
\end{split}
\end{equation}
which has the form of a finite-difference Laplacian of the
polygamma function $\psi_\alpha(z)$, defined as the $(\alpha+1)$-th
derivative of the logarithm of the gamma function $\Gamma(z)$.

\begin{figure*}[t]
\includegraphics[width=1 \textwidth]{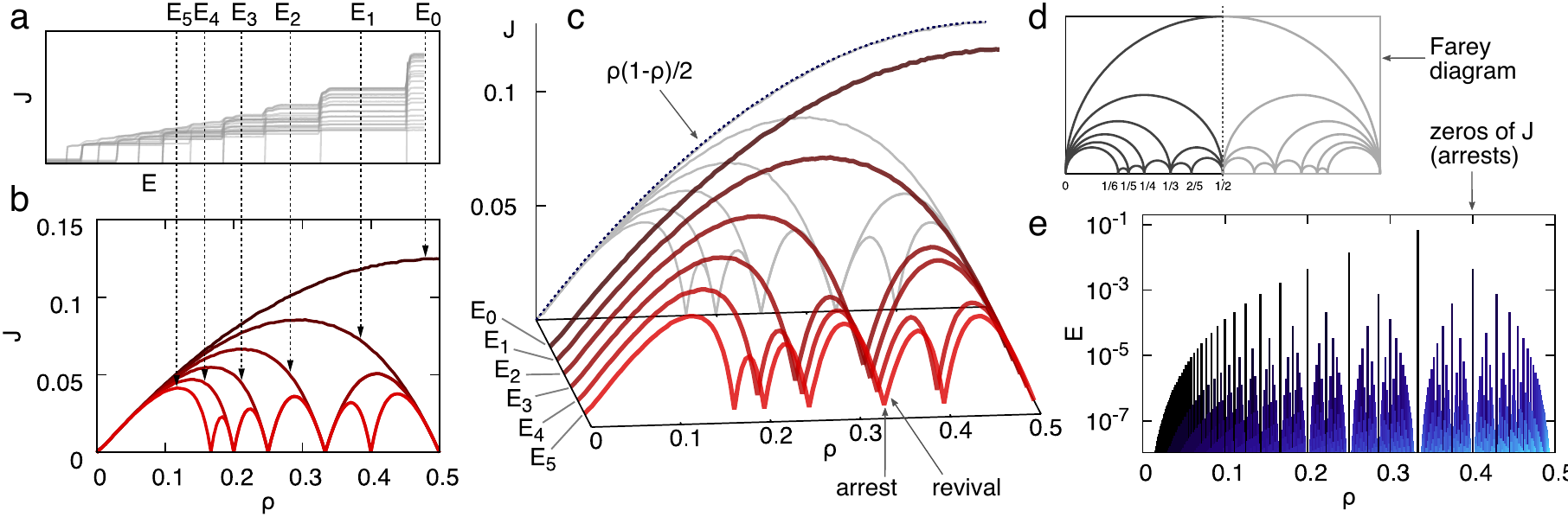}
\onecolumngrid
\caption{
The density dependence of the stationary current $J$ exposes a fractal hierarchy of excitations.
Six values of the driving field $E$, chosen as shown in panel (a),
which is the same plot as in Fig.~\ref{figure:JvsE}a,
give the red solid curves in panels (b,c).
The dotted line in (c) is the mean-field formula for large fields $E>E_\mathrm{sat}$.
The fractal structure of the current-carrying excitations
is captured by the Farey diagram
of number theory, panel (d).
As $E$ decreases, more and more densities crystallize.
Panel (e) shows all the zeros of $J(\rho)$
down to the threshold field for denominator $q=100$;
lighter colors correspond to larger numerators $p$ (black is $p=1$).
}
\label{figure:Jvsrho}
\end{figure*}

The results of this calculation are in perfect agreement with the
numerical simulations, as is shown in Fig.~\ref{figure:JvsE} for the
potential $V(x)=1/|x|^3$.  The power-law asymptotic behavior
$E_\mathrm{thr} \approx q^{-5} \left| \psi_4(1) \right|/2$, obtained
from the analytical formula, is a good approximant already from
$q=11$, with deviations below 2\%.  The asymptotic behavior in the
case $\alpha=2$ is $E_\mathrm{thr} \approx q^{-4} \, \pi^4/15$.
Due to the property (\emph{iii}) discussed above, the analytical formula
describes global features of the whole ``phase diagram'',
beyond the sole location of the breakdown transitions.
%


\subsection*{Fractal hierarchy of excitations}
At any fixed density $\rho$, the current is a non-decreasing function
of the field and reaches a maximum stationary value at large $E$.  The
saturation currents $J_\mathrm{max}(\rho)$ are those shown in
Fig.~\ref{figure:JvsE} around $E=1$.  A mean field prediction,
analogous to the standard one for simple exclusion processes,
can be obtained by assuming that interactions become irrelevant
at large fields, so that motion is constrained only by exclusion.  The
current is then proportional to the probability of finding an ordered
particle-hole pair, i.e.,
\begin{equation}
J_\mathrm{max} = \frac{\rho(1-\rho)}{2}.
\end{equation}
The mean-field formula perfectly captures the density dependence of
$J_\mathrm{max}$, as can be seen in Fig.~\ref{figure:Jvsrho}c.

The saturation field $E_\mathrm{sat}$, such that $J(\rho)=J_\mathrm{max}(\rho)$
for $E\geq E_\mathrm{sat}$,
can be calculated exactly in the
thermodynamic limit $L\to\infty$.  Owing to the convexity of the
potential, the largest energy difference in a single particle-hole
exchange is attained when the particle has an infinite number of
consecutive holes to its right and the hole has an infinite number of
particles to its left, and is therefore
$E_\mathrm{sat}=V(1)=1$.  Interestingly, the saturation field does not
depend on the density, nor on the detailed form of the potential.

By lowering the driving field from $E_\mathrm{sat}$, the density
dependence of $J$ departs from the mean-field curve.
Below $E_\mathrm{thr}(1/2)$, the system at half filling is in the
insulator phase, and the stationary current must be zero, as it is for
$\rho = 0$.  Therefore $J$ cannot be monotonic, and instead assumes a
nearly parabolic shape (Fig.~\ref{figure:Jvsrho}bc).  Below
$E_\mathrm{thr}(1/3)$, the system enters in the insulating phase for
$\rho=1/3$: the parabolic function splits into two daughter curves,
with zeroes in $\rho=\left\{0,1/2,1/3\right\}$.
In the limit of small driving fields, this process generates
iteratively a fractal hierarchy of branches.
Between $E_\mathrm{thr}(1/(q+1))$ and $E_\mathrm{thr}(1/q)$ the
current has zeros for every $\rho=p/q'$ such that $q'\leq q$ (with $p$
and $q'$ coprime).  This ordered set is called the \emph{Farey
sequence} $\mathcal F_{q'}$ of order $q'$ 
\cite{HardyWright:BOOK}.
Since these are
the only zeros, the plot of $J(\rho)$ realizes what is known as the
\emph{Farey diagram} in number theory,
constructed by connecting consecutive fractions in the
Farey sequences at all orders (see Fig.~\ref{figure:Jvsrho}d).

It is interesting to note a connection between the Farey hierarchy
described here and the fractal hierarchy of the crystalline ground
states.
It can be shown that the Hubbard state at $\rho=p/q$
can be constructed iteratively by considering the finite continued-fraction expansion
\begin{equation}
\frac{p}{q}=\cfrac{1}{a_1+\cfrac{1}{a_2+\cfrac{1}{\ddots+\cfrac{1}{a_{\lambda}}}}},
\end{equation}
which is customarily denoted by $[0;a_1,a_2,\ldots,a_\lambda]$.
At level 1, the density $1/a_1$ is simply realized by one particle followed by $a_1-1$ holes;
let us call this block $X_1$ and denote it by $(1)(0)^{a_1-1}$.
At level 2, this arrangement gets corrected, by interposing a block $Y_1=(1)(0)^{a_1}$,
which is longer and has smaller density, once every $a_2-1$ blocks of type $X_1$,
hence obtaining a new block $X_2=(Y_1)(X_1)^{a_1-1}$.
This procedure is repeated up to the final level $\lambda$:
at each level a larger-scale structure is specified
(more details can be found in \cite{Hubbard:PRB:1978}).
The foregoing construction gives rise to a hierarchy between rational densities,
whereby $\rho'\prec\rho$ ($\rho'$ precedes $\rho$) whenever the continued fraction of $\rho$
``starts with'' that of $\rho'$, meaning that $\rho'=[0;a'_1,\ldots,a'_{\lambda'}]$
and $\rho=[0;a'_1,\ldots,a'_{\lambda'},\ldots,a_{\lambda}]$.
This hierarchy is at the core of the fractal phase diagram
that arises in the equilibrium ($E=0$) model in the grand-canonical ensemble
(first studied in~\cite{Bak:PRL:1982}).
The ground-state density as a function of chemical potential in that situation
is a \emph{devil's staircase}, a self-similar Cantor function with
plateaux at every rational number;
similar fractal layouts of transitions are often found
in presence of competing interactions or scales \cite{Bak:RPP:1982, Selke:1988, AlonsoHovi:1998}.
Interestingly, there exists a one-to-one correspondence between
rational numbers and zig-zag paths on the Farey diagram, starting from
0 and alternating between rightward and leftward jumps.  This
correspondence allows to easily construct the continued fraction of a
rational number from the topology of its path (proofs of these results
can be found in \cite{Hatcher:2002,Goldman:1988}).
In this sense, the hierarchical layout of the excitations in our model
realizes the out-of-equilibrium counterpart of the hierarchy of ground
states.

Finally, we emphasize a counterintuitive aspect of $J(\rho)$,
which is surprising from the point of view of driven diffusive systems.
In classic short-range simple exclusion processes, the stationary current
can become zero at a critical density, giving rise to an arrest (a ``traffic jam''),
which persists for all larger densities.
By contrast, in the long-range DLG a jammed system can be revived by
increasing the density.
At fixed $E$, adding particles to an empty lattice of finite size
first increases the current, then decreases it to an arrest where the
system crystallizes.  At this point, adding a single particle restores
a non-zero current.
Eq.~(\ref{eqEthrPwl}) gives the complete structure of the arrests with
varying external field (Fig.~\ref{figure:Jvsrho}e) up to a much larger
detail than is attainable by simulations.


\section{Discussion}

Driven diffusive systems are useful for both conceptual and applied
reasons. On the conceptual side, they offer a broad class of
statistical physics models that do not respect detailed balance. Thus,
they provide examples of non-equilibrium stationary states,
macroscopic and microscopic currents, pattern formation and
non-equilibrium (and boundary-induced) phase transitions.  On the
applied side, they provide templates for models of specific systems,
and clear interpretation tools that are possible only with models free
of system-specific details.
We believe that the DLG with slowly-decaying repulsive interactions
defined here has potential on both the fundamental and the applied
side.

The most important conceptual contribution of the model defined here
consists in the non-trivial transport properties related to the energy
gaps between the configurations belonging to the non-equilibrium
stationary states.
The finite gap between the Hubbard ground state and the first excitation
at fixed density is responsible for the insulator breakdown,
as confirmed by the calculation of $E_\mathrm{thr}$.
This (non-equilibrium) situation resembles the equilibrium
grand-canonical case, where the crystalline ground states are
incompressible, meaning that $\kappa = \mathrm{d}\rho/\mathrm{d}\mu$
is zero in a range of chemical potentials $\mu$.
In this case, the differential conductivity $\sigma$ plays the role of
the compressibility $\kappa$.
This analogy is reinforced by the close similarity between
Eq.~(\ref{eqEthr}) and the formula giving the stability interval
$\Delta\mu$ of the ground states at density $p/q$, namely
$\Delta\mu=2 q \sum_k k [V(qk+1)+V(qk-1)-2V(qk)]$
\cite{Bak:PRL:1982}.
Notably, both $\Delta\mu$ and $E_\mathrm{thr}$ depend only on the
denominator of the particle density: this feature is responsible for
the devil's staircase phase diagram at equilibrium and for
the Farey hierarchy out of equilibrium.

On the more applied side, we believe that the repulsive driven lattice
gas may play a role in the context of hard-condensed matter, as a
simple paradigm of non-equilibrium transport showing a wealth of
anomalous behavior.
Indeed, no common interpretation tools are established for the
peculiar transitions observed in driven quantum many-body systems.
%
The correlated nature of the degrees of freedom is a major obstacle to
theoretical advances in this field.  The complex interplay between
quantum effects, interactions, and macroscopic currents, together with
the lack of natural perturbative parameters, have challenged the
traditional descriptions in terms of Fermi liquids and crippled the
development of a common theoretical framework for these
systems~\cite{Eisert:2015, JainAnderson:2009, GeorgesKotliar:1996,
  OkaArita:2003, KrinnerLebrat:2016, GeorgesKotliar:1996,
  Moreschini:2016}.
Complementarily to other studies \cite{JainAnderson:2009}, our model
suggests that repulsive interactions alone, in absence of quantum
effects, are able to produce anomalous transport phenomena that are
similar to those observed in some systems.

Whether this resemblance can be made more rigorous is an open question.  
The reduction to a classical kinetically-constrained master equation may in some cases be approached
rigorously, by starting from the quantum evolution in Lindblad
form and integrating out the fast degrees of freedom
\cite{LesanovskyGarrahan:2013}. Such approaches may reveal whether and
to what extent the model defined here may capture the physics of a
specific quantum system.
Recently, the FQHE Hamiltonian has been mapped (in the ``thin-torus''
limit) to a classical equilibrium 1D lattice gas with repulsive long-range
interactions, that is the grand-canonical version of our lattice gas at $E=0$,
whereby the lattice is realized by the quantum states in the lowest Landau level
\cite{Rotondo:PRL:2016}.
The hierarchy of quasi-particles that emerges from the continued-fraction expansion
is precisely the Haldane-Halperin hierarchy of the FQHE
\cite{BergholtzHansson:2007, Haldane:1983, Halperin:1984}.
The fixed-current plateaux in our model correspond to current-carrying
excitations, whose layout as a function of $\rho$ in the $J$ versus
$\rho$ plane (Fig.~\ref{figure:Jvsrho}) parallels the FQHE phase
diagram \cite{BergholtzHansson:2007, KivelsonLee:1992}.
Here, the excitations realise the Farey hierarchy,
somewhat complementarily to the Haldane-Halperin case.
Also, the stability of a ground state under the external field $E$
is analogous to the stability of a FQHE state
with respect to sample disorder, in the way the
stability thresholds
depend on the denominator of the density~\cite{BergholtzHansson:2007}.

In conclusion, the repulsive driven lattice gas defined here shows how
some highly non-linear transport properties, closely resembling those
in quantum many-body systems, can originate from a classical mechanism
whose key ingredient are large-scale interactions.  The crucial aspect
leading to the observed behavior is the competition between the
lattice scale and the inverse density.
We stress that the phenomenology of the model is independent of the
precise form of the potential (and thus in some sense ``universal''),
thus pointing to the importance for anomalous transport of the
conflict between interactions and system-intrinsic length scales.

\vspace{0.3cm}
\hyphenation{Ca-rac-cio-lo}
\begin{acknowledgments}
  We are greatful to Bruno Bassetti, Pietro Cicuta, Juan P.~Garrahan,
  and Luca Moreschini for useful discussions, and to Alessandro Vicini
  for providing computer time.
\end{acknowledgments}


\begin{thebibliography}{49}%
\makeatletter
\providecommand \@ifxundefined [1]{%
 \@ifx{#1\undefined}
}%
\providecommand \@ifnum [1]{%
 \ifnum #1\expandafter \@firstoftwo
 \else \expandafter \@secondoftwo
 \fi
}%
\providecommand \@ifx [1]{%
 \ifx #1\expandafter \@firstoftwo
 \else \expandafter \@secondoftwo
 \fi
}%
\providecommand \natexlab [1]{#1}%
\providecommand \enquote  [1]{``#1''}%
\providecommand \bibnamefont  [1]{#1}%
\providecommand \bibfnamefont [1]{#1}%
\providecommand \citenamefont [1]{#1}%
\providecommand \href@noop [0]{\@secondoftwo}%
\providecommand \href [0]{\begingroup \@sanitize@url \@href}%
\providecommand \@href[1]{\@@startlink{#1}\@@href}%
\providecommand \@@href[1]{\endgroup#1\@@endlink}%
\providecommand \@sanitize@url [0]{\catcode `\\12\catcode `\$12\catcode
  `\&12\catcode `\#12\catcode `\^12\catcode `\_12\catcode `\%12\relax}%
\providecommand \@@startlink[1]{}%
\providecommand \@@endlink[0]{}%
\providecommand \url  [0]{\begingroup\@sanitize@url \@url }%
\providecommand \@url [1]{\endgroup\@href {#1}{\urlprefix }}%
\providecommand \urlprefix  [0]{URL }%
\providecommand \Eprint [0]{\href }%
\providecommand \doibase [0]{http://dx.doi.org/}%
\providecommand \selectlanguage [0]{\@gobble}%
\providecommand \bibinfo  [0]{\@secondoftwo}%
\providecommand \bibfield  [0]{\@secondoftwo}%
\providecommand \translation [1]{[#1]}%
\providecommand \BibitemOpen [0]{}%
\providecommand \bibitemStop [0]{}%
\providecommand \bibitemNoStop [0]{.\EOS\space}%
\providecommand \EOS [0]{\spacefactor3000\relax}%
\providecommand \BibitemShut  [1]{\csname bibitem#1\endcsname}%
\let\auto@bib@innerbib\@empty
\bibitem [{\citenamefont {Katz}\ \emph {et~al.}(1984)\citenamefont {Katz},
  \citenamefont {Lebowitz},\ and\ \citenamefont {Spohn}}]{Katz:1984}%
  \BibitemOpen
  \bibfield  {author} {\bibinfo {author} {\bibfnamefont {S.}~\bibnamefont
  {Katz}}, \bibinfo {author} {\bibfnamefont {J.~L.}\ \bibnamefont {Lebowitz}},
  \ and\ \bibinfo {author} {\bibfnamefont {H.}~\bibnamefont {Spohn}},\ }\href
  {\doibase 10.1007/BF01018556} {\bibfield  {journal} {\bibinfo  {journal}
  {Journal of Statistical Physics}\ }\textbf {\bibinfo {volume} {34}},\
  \bibinfo {pages} {497} (\bibinfo {year} {1984})}\BibitemShut {NoStop}%
\bibitem [{\citenamefont {Schmittmann}\ and\ \citenamefont
  {Zia}(1995)}]{SchmittmannZia:1995}%
  \BibitemOpen
  \bibfield  {author} {\bibinfo {author} {\bibfnamefont {B.}~\bibnamefont
  {Schmittmann}}\ and\ \bibinfo {author} {\bibfnamefont {R.}~\bibnamefont
  {Zia}},\ }in\ \href {\doibase
  http://dx.doi.org/10.1016/S1062-7901(06)80014-5} {\emph {\bibinfo {booktitle}
  {Statistical Mechanics of Driven Diffusive System}}},\ \bibinfo {series}
  {Phase Transitions and Critical Phenomena}, Vol.~\bibinfo {volume} {17},\
  \bibinfo {editor} {edited by\ \bibinfo {editor} {\bibfnamefont
  {C.}~\bibnamefont {Domb}}\ and\ \bibinfo {editor} {\bibfnamefont {J.~L.}\
  \bibnamefont {Lebowitz}}}\ (\bibinfo  {publisher} {Academic Press, London},\
  \bibinfo {year} {1995})\ pp.\ \bibinfo {pages} {3 -- 214}\BibitemShut
  {NoStop}%
\bibitem [{\citenamefont {Chowdhury}\ \emph {et~al.}(2000)\citenamefont
  {Chowdhury}, \citenamefont {Santen},\ and\ \citenamefont
  {Schadschneider}}]{Chowdhury2000}%
  \BibitemOpen
  \bibfield  {author} {\bibinfo {author} {\bibfnamefont {D.}~\bibnamefont
  {Chowdhury}}, \bibinfo {author} {\bibfnamefont {L.}~\bibnamefont {Santen}}, \
  and\ \bibinfo {author} {\bibfnamefont {A.}~\bibnamefont {Schadschneider}},\
  }\href@noop {} {\bibfield  {journal} {\bibinfo  {journal} {Physics Reports}\
  }\textbf {\bibinfo {volume} {329}},\ \bibinfo {pages} {199} (\bibinfo {year}
  {2000})}\BibitemShut {NoStop}%
\bibitem [{\citenamefont {Chowdhury}\ \emph {et~al.}(2004)\citenamefont
  {Chowdhury}, \citenamefont {Nishinari},\ and\ \citenamefont
  {Schadschneider}}]{Chowdhury2004}%
  \BibitemOpen
  \bibfield  {author} {\bibinfo {author} {\bibfnamefont {D.}~\bibnamefont
  {Chowdhury}}, \bibinfo {author} {\bibfnamefont {K.}~\bibnamefont
  {Nishinari}}, \ and\ \bibinfo {author} {\bibfnamefont {A.}~\bibnamefont
  {Schadschneider}},\ }\href@noop {} {\bibfield  {journal} {\bibinfo  {journal}
  {Phase Transitions}\ }\textbf {\bibinfo {volume} {77}},\ \bibinfo {pages}
  {601} (\bibinfo {year} {2004})}\BibitemShut {NoStop}%
\bibitem [{\citenamefont {Chowdhury}\ \emph {et~al.}(2005)\citenamefont
  {Chowdhury}, \citenamefont {Schadschneider},\ and\ \citenamefont
  {Nishinari}}]{Chowdhury2005}%
  \BibitemOpen
  \bibfield  {author} {\bibinfo {author} {\bibfnamefont {D.}~\bibnamefont
  {Chowdhury}}, \bibinfo {author} {\bibfnamefont {A.}~\bibnamefont
  {Schadschneider}}, \ and\ \bibinfo {author} {\bibfnamefont {K.}~\bibnamefont
  {Nishinari}},\ }\href@noop {} {\bibfield  {journal} {\bibinfo  {journal}
  {Physics of Life reviews}\ }\textbf {\bibinfo {volume} {2}},\ \bibinfo
  {pages} {318} (\bibinfo {year} {2005})}\BibitemShut {NoStop}%
\bibitem [{\citenamefont {Parmeggiani}\ \emph
  {et~al.}(2004{\natexlab{a}})\citenamefont {Parmeggiani}, \citenamefont
  {Franosch},\ and\ \citenamefont {Frey}}]{ParmeggianiFranosch:2004}%
  \BibitemOpen
  \bibfield  {author} {\bibinfo {author} {\bibfnamefont {A.}~\bibnamefont
  {Parmeggiani}}, \bibinfo {author} {\bibfnamefont {T.}~\bibnamefont
  {Franosch}}, \ and\ \bibinfo {author} {\bibfnamefont {E.}~\bibnamefont
  {Frey}},\ }\href {\doibase 10.1103/PhysRevE.70.046101} {\bibfield  {journal}
  {\bibinfo  {journal} {Phys. Rev. E}\ }\textbf {\bibinfo {volume} {70}},\
  \bibinfo {pages} {046101} (\bibinfo {year} {2004}{\natexlab{a}})}\BibitemShut
  {NoStop}%
\bibitem [{\citenamefont {Parmeggiani}\ \emph
  {et~al.}(2004{\natexlab{b}})\citenamefont {Parmeggiani}, \citenamefont
  {Franosch},\ and\ \citenamefont {Frey}}]{Parmeggiani2004}%
  \BibitemOpen
  \bibfield  {author} {\bibinfo {author} {\bibfnamefont {A.}~\bibnamefont
  {Parmeggiani}}, \bibinfo {author} {\bibfnamefont {T.}~\bibnamefont
  {Franosch}}, \ and\ \bibinfo {author} {\bibfnamefont {E.}~\bibnamefont
  {Frey}},\ }\href@noop {} {\bibfield  {journal} {\bibinfo  {journal} {Physical
  Review E}\ }\textbf {\bibinfo {volume} {70}},\ \bibinfo {pages} {046101}
  (\bibinfo {year} {2004}{\natexlab{b}})}\BibitemShut {NoStop}%
\bibitem [{\citenamefont {Parmeggiani}\ \emph {et~al.}(2003)\citenamefont
  {Parmeggiani}, \citenamefont {Franosch},\ and\ \citenamefont
  {Frey}}]{Parmeggiani2003}%
  \BibitemOpen
  \bibfield  {author} {\bibinfo {author} {\bibfnamefont {A.}~\bibnamefont
  {Parmeggiani}}, \bibinfo {author} {\bibfnamefont {T.}~\bibnamefont
  {Franosch}}, \ and\ \bibinfo {author} {\bibfnamefont {E.}~\bibnamefont
  {Frey}},\ }\href@noop {} {\bibfield  {journal} {\bibinfo  {journal} {Physical
  review letters}\ }\textbf {\bibinfo {volume} {90}},\ \bibinfo {pages}
  {086601} (\bibinfo {year} {2003})}\BibitemShut {NoStop}%
\bibitem [{\citenamefont {Lipowsky}\ \emph {et~al.}(2001)\citenamefont
  {Lipowsky}, \citenamefont {Klumpp},\ and\ \citenamefont
  {Nieuwenhuizen}}]{Lipowsky2001}%
  \BibitemOpen
  \bibfield  {author} {\bibinfo {author} {\bibfnamefont {R.}~\bibnamefont
  {Lipowsky}}, \bibinfo {author} {\bibfnamefont {S.}~\bibnamefont {Klumpp}}, \
  and\ \bibinfo {author} {\bibfnamefont {T.~M.}\ \bibnamefont
  {Nieuwenhuizen}},\ }\href@noop {} {\bibfield  {journal} {\bibinfo  {journal}
  {Physical Review Letters}\ }\textbf {\bibinfo {volume} {87}},\ \bibinfo
  {pages} {108101} (\bibinfo {year} {2001})}\BibitemShut {NoStop}%
\bibitem [{\citenamefont {Becker}\ \emph {et~al.}(2013)\citenamefont {Becker},
  \citenamefont {Nelissen}, \citenamefont {Cleuren}, \citenamefont {Partoens},\
  and\ \citenamefont {Van~den Broeck}}]{BeckerNelissen:2013}%
  \BibitemOpen
  \bibfield  {author} {\bibinfo {author} {\bibfnamefont {T.}~\bibnamefont
  {Becker}}, \bibinfo {author} {\bibfnamefont {K.}~\bibnamefont {Nelissen}},
  \bibinfo {author} {\bibfnamefont {B.}~\bibnamefont {Cleuren}}, \bibinfo
  {author} {\bibfnamefont {B.}~\bibnamefont {Partoens}}, \ and\ \bibinfo
  {author} {\bibfnamefont {C.}~\bibnamefont {Van~den Broeck}},\ }\href
  {\doibase 10.1103/PhysRevLett.111.110601} {\bibfield  {journal} {\bibinfo
  {journal} {Phys. Rev. Lett.}\ }\textbf {\bibinfo {volume} {111}},\ \bibinfo
  {pages} {110601} (\bibinfo {year} {2013})}\BibitemShut {NoStop}%
\bibitem [{\citenamefont {Tailleur}\ \emph {et~al.}(2007)\citenamefont
  {Tailleur}, \citenamefont {Kurchan},\ and\ \citenamefont
  {Lecomte}}]{TailleurKurchan:2007}%
  \BibitemOpen
  \bibfield  {author} {\bibinfo {author} {\bibfnamefont {J.}~\bibnamefont
  {Tailleur}}, \bibinfo {author} {\bibfnamefont {J.}~\bibnamefont {Kurchan}}, \
  and\ \bibinfo {author} {\bibfnamefont {V.}~\bibnamefont {Lecomte}},\ }\href
  {\doibase 10.1103/PhysRevLett.99.150602} {\bibfield  {journal} {\bibinfo
  {journal} {Phys. Rev. Lett.}\ }\textbf {\bibinfo {volume} {99}},\ \bibinfo
  {pages} {150602} (\bibinfo {year} {2007})}\BibitemShut {NoStop}%
\bibitem [{\citenamefont {Caracciolo}\ \emph {et~al.}(2003)\citenamefont
  {Caracciolo}, \citenamefont {Gambassi}, \citenamefont {Gubinelli},\ and\
  \citenamefont {Pelissetto}}]{CaraccioloGambassi:2003}%
  \BibitemOpen
  \bibfield  {author} {\bibinfo {author} {\bibfnamefont {S.}~\bibnamefont
  {Caracciolo}}, \bibinfo {author} {\bibfnamefont {A.}~\bibnamefont
  {Gambassi}}, \bibinfo {author} {\bibfnamefont {M.}~\bibnamefont {Gubinelli}},
  \ and\ \bibinfo {author} {\bibfnamefont {A.}~\bibnamefont {Pelissetto}},\
  }\href {http://stacks.iop.org/0305-4470/36/i=21/a=101} {\bibfield  {journal}
  {\bibinfo  {journal} {Journal of Physics A: Mathematical and General}\
  }\textbf {\bibinfo {volume} {36}},\ \bibinfo {pages} {L315} (\bibinfo {year}
  {2003})}\BibitemShut {NoStop}%
\bibitem [{\citenamefont {de~los Santos}\ \emph {et~al.}(2001)\citenamefont
  {de~los Santos}, \citenamefont {Mu\~{n}oz},\ and\ \citenamefont
  {Garrido}}]{delosSantosMunoz:2001}%
  \BibitemOpen
  \bibfield  {author} {\bibinfo {author} {\bibfnamefont {F.}~\bibnamefont
  {de~los Santos}}, \bibinfo {author} {\bibfnamefont {M.~A.}\ \bibnamefont
  {Mu\~{n}oz}}, \ and\ \bibinfo {author} {\bibfnamefont {P.~L.}\ \bibnamefont
  {Garrido}},\ }\href {\doibase http://dx.doi.org/10.1063/1.1386833} {\bibfield
   {journal} {\bibinfo  {journal} {AIP Conference Proceedings}\ }\textbf
  {\bibinfo {volume} {574}},\ \bibinfo {pages} {149} (\bibinfo {year}
  {2001})}\BibitemShut {NoStop}%
\bibitem [{\citenamefont {Levi}\ and\ \citenamefont
  {Lesanovsky}(2014)}]{LeviLesanovsky:2014}%
  \BibitemOpen
  \bibfield  {author} {\bibinfo {author} {\bibfnamefont {E.}~\bibnamefont
  {Levi}}\ and\ \bibinfo {author} {\bibfnamefont {I.}~\bibnamefont
  {Lesanovsky}},\ }\href {http://stacks.iop.org/1367-2630/16/i=9/a=093053}
  {\bibfield  {journal} {\bibinfo  {journal} {New Journal of Physics}\ }\textbf
  {\bibinfo {volume} {16}},\ \bibinfo {pages} {093053} (\bibinfo {year}
  {2014})}\BibitemShut {NoStop}%
\bibitem [{\citenamefont {Olmos}\ \emph {et~al.}(2013)\citenamefont {Olmos},
  \citenamefont {Yu}, \citenamefont {Singh}, \citenamefont {Schreck},
  \citenamefont {Bongs},\ and\ \citenamefont {Lesanovsky}}]{OlmosYu:2013}%
  \BibitemOpen
  \bibfield  {author} {\bibinfo {author} {\bibfnamefont {B.}~\bibnamefont
  {Olmos}}, \bibinfo {author} {\bibfnamefont {D.}~\bibnamefont {Yu}}, \bibinfo
  {author} {\bibfnamefont {Y.}~\bibnamefont {Singh}}, \bibinfo {author}
  {\bibfnamefont {F.}~\bibnamefont {Schreck}}, \bibinfo {author} {\bibfnamefont
  {K.}~\bibnamefont {Bongs}}, \ and\ \bibinfo {author} {\bibfnamefont
  {I.}~\bibnamefont {Lesanovsky}},\ }\href {\doibase
  10.1103/PhysRevLett.110.143602} {\bibfield  {journal} {\bibinfo  {journal}
  {Phys. Rev. Lett.}\ }\textbf {\bibinfo {volume} {110}},\ \bibinfo {pages}
  {143602} (\bibinfo {year} {2013})}\BibitemShut {NoStop}%
\bibitem [{\citenamefont {Rademaker}\ \emph {et~al.}(2013)\citenamefont
  {Rademaker}, \citenamefont {Pramudya}, \citenamefont {Zaanen},\ and\
  \citenamefont {Dobrosavljevi\ifmmode~\acute{c}\else
  \'{c}\fi{}}}]{Rademaker:2013}%
  \BibitemOpen
  \bibfield  {author} {\bibinfo {author} {\bibfnamefont {L.}~\bibnamefont
  {Rademaker}}, \bibinfo {author} {\bibfnamefont {Y.}~\bibnamefont {Pramudya}},
  \bibinfo {author} {\bibfnamefont {J.}~\bibnamefont {Zaanen}}, \ and\ \bibinfo
  {author} {\bibfnamefont {V.}~\bibnamefont
  {Dobrosavljevi\ifmmode~\acute{c}\else \'{c}\fi{}}},\ }\href {\doibase
  10.1103/PhysRevE.88.032121} {\bibfield  {journal} {\bibinfo  {journal} {Phys.
  Rev. E}\ }\textbf {\bibinfo {volume} {88}},\ \bibinfo {pages} {032121}
  (\bibinfo {year} {2013})}\BibitemShut {NoStop}%
\bibitem [{\citenamefont {Za\l{}uska-Kotur}\ and\ \citenamefont
  {Gortel}(2006)}]{SaluskaKotur:2006}%
  \BibitemOpen
  \bibfield  {author} {\bibinfo {author} {\bibfnamefont {M.~A.}\ \bibnamefont
  {Za\l{}uska-Kotur}}\ and\ \bibinfo {author} {\bibfnamefont {Z.~W.}\
  \bibnamefont {Gortel}},\ }\href {\doibase 10.1103/PhysRevB.74.045405}
  {\bibfield  {journal} {\bibinfo  {journal} {Phys. Rev. B}\ }\textbf {\bibinfo
  {volume} {74}},\ \bibinfo {pages} {045405} (\bibinfo {year}
  {2006})}\BibitemShut {NoStop}%
\bibitem [{\citenamefont {Schau{\ss}}\ \emph {et~al.}(2015)\citenamefont
  {Schau{\ss}}, \citenamefont {Zeiher}, \citenamefont {Fukuhara}, \citenamefont
  {Hild}, \citenamefont {Cheneau}, \citenamefont {Macr{\`\i}}, \citenamefont
  {Pohl}, \citenamefont {Bloch},\ and\ \citenamefont
  {Gross}}]{Schauss:Science:2015}%
  \BibitemOpen
  \bibfield  {author} {\bibinfo {author} {\bibfnamefont {P.}~\bibnamefont
  {Schau{\ss}}}, \bibinfo {author} {\bibfnamefont {J.}~\bibnamefont {Zeiher}},
  \bibinfo {author} {\bibfnamefont {T.}~\bibnamefont {Fukuhara}}, \bibinfo
  {author} {\bibfnamefont {S.}~\bibnamefont {Hild}}, \bibinfo {author}
  {\bibfnamefont {M.}~\bibnamefont {Cheneau}}, \bibinfo {author} {\bibfnamefont
  {T.}~\bibnamefont {Macr{\`\i}}}, \bibinfo {author} {\bibfnamefont
  {T.}~\bibnamefont {Pohl}}, \bibinfo {author} {\bibfnamefont {I.}~\bibnamefont
  {Bloch}}, \ and\ \bibinfo {author} {\bibfnamefont {C.}~\bibnamefont
  {Gross}},\ }\href {\doibase 10.1126/science.1258351} {\bibfield  {journal}
  {\bibinfo  {journal} {Science}\ }\textbf {\bibinfo {volume} {347}},\ \bibinfo
  {pages} {1455} (\bibinfo {year} {2015})}\BibitemShut {NoStop}%
\bibitem [{\citenamefont {Zhang}\ \emph {et~al.}(2015)\citenamefont {Zhang},
  \citenamefont {Fan}, \citenamefont {Liang}, \citenamefont {Ma}, \citenamefont
  {Chen}, \citenamefont {Jia},\ and\ \citenamefont {Nori}}]{ZhangFan:2015}%
  \BibitemOpen
  \bibfield  {author} {\bibinfo {author} {\bibfnamefont {Y.}~\bibnamefont
  {Zhang}}, \bibinfo {author} {\bibfnamefont {J.}~\bibnamefont {Fan}}, \bibinfo
  {author} {\bibfnamefont {J.~Q.}\ \bibnamefont {Liang}}, \bibinfo {author}
  {\bibfnamefont {J.}~\bibnamefont {Ma}}, \bibinfo {author} {\bibfnamefont
  {G.}~\bibnamefont {Chen}}, \bibinfo {author} {\bibfnamefont {S.}~\bibnamefont
  {Jia}}, \ and\ \bibinfo {author} {\bibfnamefont {F.}~\bibnamefont {Nori}},\
  }\href {http://dx.doi.org/10.1038/srep11510} {\ \textbf {\bibinfo {volume}
  {5}},\ \bibinfo {pages} {11510 EP } (\bibinfo {year} {2015})}\BibitemShut
  {NoStop}%
\bibitem [{\citenamefont {Lauer}\ \emph {et~al.}(2012)\citenamefont {Lauer},
  \citenamefont {Muth},\ and\ \citenamefont {Fleischhauer}}]{LauerMuth:2012}%
  \BibitemOpen
  \bibfield  {author} {\bibinfo {author} {\bibfnamefont {A.}~\bibnamefont
  {Lauer}}, \bibinfo {author} {\bibfnamefont {D.}~\bibnamefont {Muth}}, \ and\
  \bibinfo {author} {\bibfnamefont {M.}~\bibnamefont {Fleischhauer}},\ }\href
  {http://stacks.iop.org/1367-2630/14/i=9/a=095009} {\bibfield  {journal}
  {\bibinfo  {journal} {New Journal of Physics}\ }\textbf {\bibinfo {volume}
  {14}},\ \bibinfo {pages} {095009} (\bibinfo {year} {2012})}\BibitemShut
  {NoStop}%
\bibitem [{\citenamefont {Hubbard}(1978)}]{Hubbard:PRB:1978}%
  \BibitemOpen
  \bibfield  {author} {\bibinfo {author} {\bibfnamefont {J.}~\bibnamefont
  {Hubbard}},\ }\href@noop {} {\bibfield  {journal} {\bibinfo  {journal} {Phys.
  Rev. B}\ }\textbf {\bibinfo {volume} {17}},\ \bibinfo {pages} {494} (\bibinfo
  {year} {1978})}\BibitemShut {NoStop}%
\bibitem [{\citenamefont {Bak}\ and\ \citenamefont
  {Bruinsma}(1982)}]{Bak:PRL:1982}%
  \BibitemOpen
  \bibfield  {author} {\bibinfo {author} {\bibfnamefont {P.}~\bibnamefont
  {Bak}}\ and\ \bibinfo {author} {\bibfnamefont {R.}~\bibnamefont {Bruinsma}},\
  }\href@noop {} {\bibfield  {journal} {\bibinfo  {journal} {Phys. Rev. Lett.}\
  }\textbf {\bibinfo {volume} {49}},\ \bibinfo {pages} {249} (\bibinfo {year}
  {1982})}\BibitemShut {NoStop}%
\bibitem [{\citenamefont {Bak}(1982)}]{Bak:RPP:1982}%
  \BibitemOpen
  \bibfield  {author} {\bibinfo {author} {\bibfnamefont {P.}~\bibnamefont
  {Bak}},\ }\href@noop {} {\bibfield  {journal} {\bibinfo  {journal} {Reports
  on Progress in Physics}\ }\textbf {\bibinfo {volume} {45}},\ \bibinfo {pages}
  {587} (\bibinfo {year} {1982})}\BibitemShut {NoStop}%
\bibitem [{\citenamefont {L\"ow}\ \emph {et~al.}(1994)\citenamefont {L\"ow},
  \citenamefont {Emery}, \citenamefont {Fabricius},\ and\ \citenamefont
  {Kivelson}}]{LowEmery:1994}%
  \BibitemOpen
  \bibfield  {author} {\bibinfo {author} {\bibfnamefont {U.}~\bibnamefont
  {L\"ow}}, \bibinfo {author} {\bibfnamefont {V.~J.}\ \bibnamefont {Emery}},
  \bibinfo {author} {\bibfnamefont {K.}~\bibnamefont {Fabricius}}, \ and\
  \bibinfo {author} {\bibfnamefont {S.~A.}\ \bibnamefont {Kivelson}},\ }\href
  {\doibase 10.1103/PhysRevLett.72.1918} {\bibfield  {journal} {\bibinfo
  {journal} {Phys. Rev. Lett.}\ }\textbf {\bibinfo {volume} {72}},\ \bibinfo
  {pages} {1918} (\bibinfo {year} {1994})}\BibitemShut {NoStop}%
\bibitem [{\citenamefont {Braun}\ and\ \citenamefont
  {Kivshar}(2004)}]{BraunKivshar:BOOK}%
  \BibitemOpen
  \bibfield  {author} {\bibinfo {author} {\bibfnamefont {O.}~\bibnamefont
  {Braun}}\ and\ \bibinfo {author} {\bibfnamefont {Y.}~\bibnamefont
  {Kivshar}},\ }\href {https://books.google.it/books?id=zyoT068mu0YC} {\emph
  {\bibinfo {title} {The Frenkel-Kontorova Model: Concepts, Methods, and
  Applications}}},\ Physics and Astronomy Online Library\ (\bibinfo
  {publisher} {Springer},\ \bibinfo {year} {2004})\BibitemShut {NoStop}%
\bibitem [{\citenamefont {Selke}(1988)}]{Selke:1988}%
  \BibitemOpen
  \bibfield  {author} {\bibinfo {author} {\bibfnamefont {W.}~\bibnamefont
  {Selke}},\ }\href {\doibase http://dx.doi.org/10.1016/0370-1573(88)90140-8}
  {\bibfield  {journal} {\bibinfo  {journal} {Physics Reports}\ }\textbf
  {\bibinfo {volume} {170}},\ \bibinfo {pages} {213 } (\bibinfo {year}
  {1988})}\BibitemShut {NoStop}%
\bibitem [{\citenamefont {Dagotto}(2005)}]{Dagotto:2005}%
  \BibitemOpen
  \bibfield  {author} {\bibinfo {author} {\bibfnamefont {E.}~\bibnamefont
  {Dagotto}},\ }\href {\doibase 10.1126/science.1107559} {\bibfield  {journal}
  {\bibinfo  {journal} {Science}\ }\textbf {\bibinfo {volume} {309}},\ \bibinfo
  {pages} {257} (\bibinfo {year} {2005})}\BibitemShut {NoStop}%
\bibitem [{\citenamefont {Nakamura}\ \emph {et~al.}(2013)\citenamefont
  {Nakamura}, \citenamefont {Sakaki}, \citenamefont {Yamanaka}, \citenamefont
  {Tamaru}, \citenamefont {Suzuki},\ and\ \citenamefont
  {Maeno}}]{Nakamura:2013}%
  \BibitemOpen
  \bibfield  {author} {\bibinfo {author} {\bibfnamefont {F.}~\bibnamefont
  {Nakamura}}, \bibinfo {author} {\bibfnamefont {M.}~\bibnamefont {Sakaki}},
  \bibinfo {author} {\bibfnamefont {Y.}~\bibnamefont {Yamanaka}}, \bibinfo
  {author} {\bibfnamefont {S.}~\bibnamefont {Tamaru}}, \bibinfo {author}
  {\bibfnamefont {T.}~\bibnamefont {Suzuki}}, \ and\ \bibinfo {author}
  {\bibfnamefont {Y.}~\bibnamefont {Maeno}},\ }\href
  {http://dx.doi.org/10.1038/srep02536} {\bibfield  {journal} {\bibinfo
  {journal} {Scientific Reports}\ }\textbf {\bibinfo {volume} {3}},\ \bibinfo
  {pages} {2536 EP } (\bibinfo {year} {2013})}\BibitemShut {NoStop}%
\bibitem [{\citenamefont {Zybtsev}\ \emph {et~al.}(2010)\citenamefont
  {Zybtsev}, \citenamefont {Pokrovskii},\ and\ \citenamefont
  {Zaitsev-Zotov}}]{Zybtsev:2010}%
  \BibitemOpen
  \bibfield  {author} {\bibinfo {author} {\bibfnamefont {S.~G.}\ \bibnamefont
  {Zybtsev}}, \bibinfo {author} {\bibfnamefont {V.~Y.}\ \bibnamefont
  {Pokrovskii}}, \ and\ \bibinfo {author} {\bibfnamefont {S.~V.}\ \bibnamefont
  {Zaitsev-Zotov}},\ }\href {http://dx.doi.org/10.1038/ncomms1087} {\bibfield
  {journal} {\bibinfo  {journal} {Nature Communications}\ }\textbf {\bibinfo
  {volume} {1}},\ \bibinfo {pages} {85 EP } (\bibinfo {year}
  {2010})}\BibitemShut {NoStop}%
\bibitem [{\citenamefont {Cario}\ \emph {et~al.}(2010)\citenamefont {Cario},
  \citenamefont {Vaju}, \citenamefont {Corraze}, \citenamefont {Guiot},\ and\
  \citenamefont {Janod}}]{CarioVaju:2010}%
  \BibitemOpen
  \bibfield  {author} {\bibinfo {author} {\bibfnamefont {L.}~\bibnamefont
  {Cario}}, \bibinfo {author} {\bibfnamefont {C.}~\bibnamefont {Vaju}},
  \bibinfo {author} {\bibfnamefont {B.}~\bibnamefont {Corraze}}, \bibinfo
  {author} {\bibfnamefont {V.}~\bibnamefont {Guiot}}, \ and\ \bibinfo {author}
  {\bibfnamefont {E.}~\bibnamefont {Janod}},\ }\href {\doibase
  10.1002/adma.201002521} {\bibfield  {journal} {\bibinfo  {journal} {Advanced
  Materials}\ }\textbf {\bibinfo {volume} {22}},\ \bibinfo {pages} {5193}
  (\bibinfo {year} {2010})}\BibitemShut {NoStop}%
\bibitem [{\citenamefont {Vinokur}\ \emph {et~al.}(2008)\citenamefont
  {Vinokur}, \citenamefont {Baturina}, \citenamefont {Fistul}, \citenamefont
  {Mironov}, \citenamefont {Baklanov},\ and\ \citenamefont
  {Strunk}}]{Vinokur:2008}%
  \BibitemOpen
  \bibfield  {author} {\bibinfo {author} {\bibfnamefont {V.~M.}\ \bibnamefont
  {Vinokur}}, \bibinfo {author} {\bibfnamefont {T.~I.}\ \bibnamefont
  {Baturina}}, \bibinfo {author} {\bibfnamefont {M.~V.}\ \bibnamefont
  {Fistul}}, \bibinfo {author} {\bibfnamefont {A.~Y.}\ \bibnamefont {Mironov}},
  \bibinfo {author} {\bibfnamefont {M.~R.}\ \bibnamefont {Baklanov}}, \ and\
  \bibinfo {author} {\bibfnamefont {C.}~\bibnamefont {Strunk}},\ }\href
  {http://dx.doi.org/10.1038/nature06837} {\bibfield  {journal} {\bibinfo
  {journal} {Nature}\ }\textbf {\bibinfo {volume} {452}},\ \bibinfo {pages}
  {613} (\bibinfo {year} {2008})}\BibitemShut {NoStop}%
\bibitem [{\citenamefont {Altshuler}\ \emph {et~al.}(2009)\citenamefont
  {Altshuler}, \citenamefont {Kravtsov}, \citenamefont {Lerner},\ and\
  \citenamefont {Aleiner}}]{Altshuler:PRL:2009}%
  \BibitemOpen
  \bibfield  {author} {\bibinfo {author} {\bibfnamefont {B.~L.}\ \bibnamefont
  {Altshuler}}, \bibinfo {author} {\bibfnamefont {V.~E.}\ \bibnamefont
  {Kravtsov}}, \bibinfo {author} {\bibfnamefont {I.~V.}\ \bibnamefont
  {Lerner}}, \ and\ \bibinfo {author} {\bibfnamefont {I.~L.}\ \bibnamefont
  {Aleiner}},\ }\href {\doibase 10.1103/PhysRevLett.102.176803} {\bibfield
  {journal} {\bibinfo  {journal} {Phys. Rev. Lett.}\ }\textbf {\bibinfo
  {volume} {102}},\ \bibinfo {pages} {176803} (\bibinfo {year}
  {2009})}\BibitemShut {NoStop}%
\bibitem [{\citenamefont {Helbing}\ \emph {et~al.}(1999)\citenamefont
  {Helbing}, \citenamefont {Mukamel},\ and\ \citenamefont
  {Sch\"utz}}]{Helbing:PRL:1999}%
  \BibitemOpen
  \bibfield  {author} {\bibinfo {author} {\bibfnamefont {D.}~\bibnamefont
  {Helbing}}, \bibinfo {author} {\bibfnamefont {D.}~\bibnamefont {Mukamel}}, \
  and\ \bibinfo {author} {\bibfnamefont {G.~M.}\ \bibnamefont {Sch\"utz}},\
  }\href {\doibase 10.1103/PhysRevLett.82.10} {\bibfield  {journal} {\bibinfo
  {journal} {Phys. Rev. Lett.}\ }\textbf {\bibinfo {volume} {82}},\ \bibinfo
  {pages} {10} (\bibinfo {year} {1999})}\BibitemShut {NoStop}%
\bibitem [{\citenamefont {Hardy}\ and\ \citenamefont
  {Wright}(1979)}]{HardyWright:BOOK}%
  \BibitemOpen
  \bibfield  {author} {\bibinfo {author} {\bibfnamefont {G.~H.}\ \bibnamefont
  {Hardy}}\ and\ \bibinfo {author} {\bibfnamefont {E.~M.}\ \bibnamefont
  {Wright}},\ }\href@noop {} {\emph {\bibinfo {title} {An introduction to the
  theory of numbers}}},\ Oxford Science Publications\ (\bibinfo  {publisher}
  {Clarendon Press},\ \bibinfo {address} {Oxford},\ \bibinfo {year}
  {1979})\BibitemShut {NoStop}%
\bibitem [{\citenamefont {Alonso}\ \emph {et~al.}(1998)\citenamefont {Alonso},
  \citenamefont {Hovi},\ and\ \citenamefont {Herrmann}}]{AlonsoHovi:1998}%
  \BibitemOpen
  \bibfield  {author} {\bibinfo {author} {\bibfnamefont {J.~J.}\ \bibnamefont
  {Alonso}}, \bibinfo {author} {\bibfnamefont {J.-P.}\ \bibnamefont {Hovi}}, \
  and\ \bibinfo {author} {\bibfnamefont {H.~J.}\ \bibnamefont {Herrmann}},\
  }\href {\doibase 10.1103/PhysRevE.58.672} {\bibfield  {journal} {\bibinfo
  {journal} {Phys. Rev. E}\ }\textbf {\bibinfo {volume} {58}},\ \bibinfo
  {pages} {672} (\bibinfo {year} {1998})}\BibitemShut {NoStop}%
\bibitem [{\citenamefont {Hatcher}(2002)}]{Hatcher:2002}%
  \BibitemOpen
  \bibfield  {author} {\bibinfo {author} {\bibfnamefont {A.}~\bibnamefont
  {Hatcher}},\ }\href@noop {} {\enquote {\bibinfo {title} {Topology of
  numbers},}\ } (\bibinfo {year} {2002}),\ \bibinfo {note}
  {http://www.math.cornell.edu/~hatcher/ (accessed 2 Nov 2016)}\BibitemShut
  {NoStop}%
\bibitem [{\citenamefont {Goldman}(1988)}]{Goldman:1988}%
  \BibitemOpen
  \bibfield  {author} {\bibinfo {author} {\bibfnamefont {J.~R.}\ \bibnamefont
  {Goldman}},\ }\href {\doibase http://dx.doi.org/10.1016/0001-8708(88)90029-1}
  {\bibfield  {journal} {\bibinfo  {journal} {Advances in Mathematics}\
  }\textbf {\bibinfo {volume} {72}},\ \bibinfo {pages} {239 } (\bibinfo {year}
  {1988})}\BibitemShut {NoStop}%
\bibitem [{\citenamefont {Eisert}\ \emph {et~al.}(2015)\citenamefont {Eisert},
  \citenamefont {Friesdorf},\ and\ \citenamefont {Gogolin}}]{Eisert:2015}%
  \BibitemOpen
  \bibfield  {author} {\bibinfo {author} {\bibfnamefont {J.}~\bibnamefont
  {Eisert}}, \bibinfo {author} {\bibfnamefont {M.}~\bibnamefont {Friesdorf}}, \
  and\ \bibinfo {author} {\bibfnamefont {C.}~\bibnamefont {Gogolin}},\ }\href
  {http://dx.doi.org/10.1038/nphys3215} {\bibfield  {journal} {\bibinfo
  {journal} {Nat Phys}\ }\textbf {\bibinfo {volume} {11}},\ \bibinfo {pages}
  {124} (\bibinfo {year} {2015})}\BibitemShut {NoStop}%
\bibitem [{\citenamefont {Jain}\ and\ \citenamefont
  {Anderson}(2009)}]{JainAnderson:2009}%
  \BibitemOpen
  \bibfield  {author} {\bibinfo {author} {\bibfnamefont {J.}~\bibnamefont
  {Jain}}\ and\ \bibinfo {author} {\bibfnamefont {P.}~\bibnamefont
  {Anderson}},\ }\href {\doibase 10.1073/pnas.0902901106} {\bibfield  {journal}
  {\bibinfo  {journal} {Proceedings of the National Academy of Sciences of the
  United States of America}\ }\textbf {\bibinfo {volume} {106}},\ \bibinfo
  {pages} {9131} (\bibinfo {year} {2009})}\BibitemShut {NoStop}%
\bibitem [{\citenamefont {Georges}\ \emph {et~al.}(1996)\citenamefont
  {Georges}, \citenamefont {Kotliar}, \citenamefont {Krauth},\ and\
  \citenamefont {Rozenberg}}]{GeorgesKotliar:1996}%
  \BibitemOpen
  \bibfield  {author} {\bibinfo {author} {\bibfnamefont {A.}~\bibnamefont
  {Georges}}, \bibinfo {author} {\bibfnamefont {G.}~\bibnamefont {Kotliar}},
  \bibinfo {author} {\bibfnamefont {W.}~\bibnamefont {Krauth}}, \ and\ \bibinfo
  {author} {\bibfnamefont {M.~J.}\ \bibnamefont {Rozenberg}},\ }\href {\doibase
  10.1103/RevModPhys.68.13} {\bibfield  {journal} {\bibinfo  {journal} {Rev.
  Mod. Phys.}\ }\textbf {\bibinfo {volume} {68}},\ \bibinfo {pages} {13}
  (\bibinfo {year} {1996})}\BibitemShut {NoStop}%
\bibitem [{\citenamefont {Oka}\ \emph {et~al.}(2003)\citenamefont {Oka},
  \citenamefont {Arita},\ and\ \citenamefont {Aoki}}]{OkaArita:2003}%
  \BibitemOpen
  \bibfield  {author} {\bibinfo {author} {\bibfnamefont {T.}~\bibnamefont
  {Oka}}, \bibinfo {author} {\bibfnamefont {R.}~\bibnamefont {Arita}}, \ and\
  \bibinfo {author} {\bibfnamefont {H.}~\bibnamefont {Aoki}},\ }\href {\doibase
  10.1103/PhysRevLett.91.066406} {\bibfield  {journal} {\bibinfo  {journal}
  {Phys. Rev. Lett.}\ }\textbf {\bibinfo {volume} {91}},\ \bibinfo {pages}
  {066406} (\bibinfo {year} {2003})}\BibitemShut {NoStop}%
\bibitem [{\citenamefont {Krinner}\ \emph {et~al.}(2016)\citenamefont
  {Krinner}, \citenamefont {Lebrat}, \citenamefont {Husmann}, \citenamefont
  {Grenier}, \citenamefont {Brantut},\ and\ \citenamefont
  {Esslinger}}]{KrinnerLebrat:2016}%
  \BibitemOpen
  \bibfield  {author} {\bibinfo {author} {\bibfnamefont {S.}~\bibnamefont
  {Krinner}}, \bibinfo {author} {\bibfnamefont {M.}~\bibnamefont {Lebrat}},
  \bibinfo {author} {\bibfnamefont {D.}~\bibnamefont {Husmann}}, \bibinfo
  {author} {\bibfnamefont {C.}~\bibnamefont {Grenier}}, \bibinfo {author}
  {\bibfnamefont {J.-P.}\ \bibnamefont {Brantut}}, \ and\ \bibinfo {author}
  {\bibfnamefont {T.}~\bibnamefont {Esslinger}},\ }\href {\doibase
  10.1073/pnas.1601812113} {\bibfield  {journal} {\bibinfo  {journal}
  {Proceedings of the National Academy of Sciences}\ }\textbf {\bibinfo
  {volume} {113}},\ \bibinfo {pages} {8144} (\bibinfo {year}
  {2016})}\BibitemShut {NoStop}%
\bibitem [{\citenamefont {Moreschini}\ \emph {et~al.}(2016)\citenamefont
  {Moreschini}, \citenamefont {Johannsen}, \citenamefont {Berger},
  \citenamefont {Denlinger}, \citenamefont {Jozwiak}, \citenamefont
  {Rotenberg}, \citenamefont {Kim}, \citenamefont {Bostwick},\ and\
  \citenamefont {Grioni}}]{Moreschini:2016}%
  \BibitemOpen
  \bibfield  {author} {\bibinfo {author} {\bibfnamefont {L.}~\bibnamefont
  {Moreschini}}, \bibinfo {author} {\bibfnamefont {J.~C.}\ \bibnamefont
  {Johannsen}}, \bibinfo {author} {\bibfnamefont {H.}~\bibnamefont {Berger}},
  \bibinfo {author} {\bibfnamefont {J.}~\bibnamefont {Denlinger}}, \bibinfo
  {author} {\bibfnamefont {C.}~\bibnamefont {Jozwiak}}, \bibinfo {author}
  {\bibfnamefont {E.}~\bibnamefont {Rotenberg}}, \bibinfo {author}
  {\bibfnamefont {K.~S.}\ \bibnamefont {Kim}}, \bibinfo {author} {\bibfnamefont
  {A.}~\bibnamefont {Bostwick}}, \ and\ \bibinfo {author} {\bibfnamefont
  {M.}~\bibnamefont {Grioni}},\ }\href {\doibase 10.1103/PhysRevB.94.081101}
  {\bibfield  {journal} {\bibinfo  {journal} {Phys. Rev. B}\ }\textbf {\bibinfo
  {volume} {94}},\ \bibinfo {pages} {081101} (\bibinfo {year}
  {2016})}\BibitemShut {NoStop}%
\bibitem [{\citenamefont {Lesanovsky}\ and\ \citenamefont
  {Garrahan}(2013)}]{LesanovskyGarrahan:2013}%
  \BibitemOpen
  \bibfield  {author} {\bibinfo {author} {\bibfnamefont {I.}~\bibnamefont
  {Lesanovsky}}\ and\ \bibinfo {author} {\bibfnamefont {J.~P.}\ \bibnamefont
  {Garrahan}},\ }\href {\doibase 10.1103/PhysRevLett.111.215305} {\bibfield
  {journal} {\bibinfo  {journal} {Phys. Rev. Lett.}\ }\textbf {\bibinfo
  {volume} {111}},\ \bibinfo {pages} {215305} (\bibinfo {year}
  {2013})}\BibitemShut {NoStop}%
\bibitem [{\citenamefont {Rotondo}\ \emph {et~al.}(2016)\citenamefont
  {Rotondo}, \citenamefont {Molinari}, \citenamefont {Ratti},\ and\
  \citenamefont {Gherardi}}]{Rotondo:PRL:2016}%
  \BibitemOpen
  \bibfield  {author} {\bibinfo {author} {\bibfnamefont {P.}~\bibnamefont
  {Rotondo}}, \bibinfo {author} {\bibfnamefont {L.~G.}\ \bibnamefont
  {Molinari}}, \bibinfo {author} {\bibfnamefont {P.}~\bibnamefont {Ratti}}, \
  and\ \bibinfo {author} {\bibfnamefont {M.}~\bibnamefont {Gherardi}},\ }\href
  {\doibase 10.1103/PhysRevLett.116.256803} {\bibfield  {journal} {\bibinfo
  {journal} {Phys. Rev. Lett.}\ }\textbf {\bibinfo {volume} {116}},\ \bibinfo
  {pages} {256803} (\bibinfo {year} {2016})}\BibitemShut {NoStop}%
\bibitem [{\citenamefont {Bergholtz}\ \emph {et~al.}(2007)\citenamefont
  {Bergholtz}, \citenamefont {Hansson}, \citenamefont {Hermanns},\ and\
  \citenamefont {Karlhede}}]{BergholtzHansson:2007}%
  \BibitemOpen
  \bibfield  {author} {\bibinfo {author} {\bibfnamefont {E.~J.}\ \bibnamefont
  {Bergholtz}}, \bibinfo {author} {\bibfnamefont {T.~H.}\ \bibnamefont
  {Hansson}}, \bibinfo {author} {\bibfnamefont {M.}~\bibnamefont {Hermanns}}, \
  and\ \bibinfo {author} {\bibfnamefont {A.}~\bibnamefont {Karlhede}},\ }\href
  {\doibase 10.1103/PhysRevLett.99.256803} {\bibfield  {journal} {\bibinfo
  {journal} {Phys. Rev. Lett.}\ }\textbf {\bibinfo {volume} {99}},\ \bibinfo
  {pages} {256803} (\bibinfo {year} {2007})}\BibitemShut {NoStop}%
\bibitem [{\citenamefont {Haldane}(1983)}]{Haldane:1983}%
  \BibitemOpen
  \bibfield  {author} {\bibinfo {author} {\bibfnamefont {F.~D.~M.}\
  \bibnamefont {Haldane}},\ }\href {\doibase 10.1103/PhysRevLett.51.605}
  {\bibfield  {journal} {\bibinfo  {journal} {Phys. Rev. Lett.}\ }\textbf
  {\bibinfo {volume} {51}},\ \bibinfo {pages} {605} (\bibinfo {year}
  {1983})}\BibitemShut {NoStop}%
\bibitem [{\citenamefont {Halperin}(1984)}]{Halperin:1984}%
  \BibitemOpen
  \bibfield  {author} {\bibinfo {author} {\bibfnamefont {B.~I.}\ \bibnamefont
  {Halperin}},\ }\href {\doibase 10.1103/PhysRevLett.52.1583} {\bibfield
  {journal} {\bibinfo  {journal} {Phys. Rev. Lett.}\ }\textbf {\bibinfo
  {volume} {52}},\ \bibinfo {pages} {1583} (\bibinfo {year}
  {1984})}\BibitemShut {NoStop}%
\bibitem [{\citenamefont {Kivelson}\ \emph {et~al.}(1992)\citenamefont
  {Kivelson}, \citenamefont {Lee},\ and\ \citenamefont
  {Zhang}}]{KivelsonLee:1992}%
  \BibitemOpen
  \bibfield  {author} {\bibinfo {author} {\bibfnamefont {S.}~\bibnamefont
  {Kivelson}}, \bibinfo {author} {\bibfnamefont {D.-H.}\ \bibnamefont {Lee}}, \
  and\ \bibinfo {author} {\bibfnamefont {S.-C.}\ \bibnamefont {Zhang}},\ }\href
  {\doibase 10.1103/PhysRevB.46.2223} {\bibfield  {journal} {\bibinfo
  {journal} {Phys. Rev. B}\ }\textbf {\bibinfo {volume} {46}},\ \bibinfo
  {pages} {2223} (\bibinfo {year} {1992})}\BibitemShut {NoStop}%
\end{thebibliography}
\end{document}